\documentstyle[preprint,aps]{revtex}
\begin{document}
\draft
\preprint{\vbox{\hbox{KWUTP-96/1}}}

\title{On Statistical Mechanics of Non-Abelian Chern-Simons Particles}
\author{Taejin Lee}

\address{Department of Physics, Kangwon National University, 
                      Chuncheon 200-701, Korea} 
%\date{}
\maketitle
\begin{abstract}
Lee replies to the comment on "Statistical Mechanics
of Non-Abelian Chern-Simons Particles" by C. R. Hage
\end{abstract}
\pacs{PACS number(s):}
\narrowtext

{\bf Lee Replies:} I recently evaluated\cite{smna} the second virial 
coefficient for the two-dimensional gas of non-Abelian Chern-Simons 
particles (NACS) \cite{nacs} which obey the non-Abelian braid statistics 
and showed that the coefficient is not periodic in the induced spin.  
In a comment on the letter Hagen\cite{hagen} claimed that the virial 
coefficient obtained in Ref.\cite{smna} is incorrect and there exists 
a periodicity in the flux parameter as in the Abelian theory\cite{arovas}. 
Here I show that the additional factor, which is claimed to be omitted 
in Eq.(28) of Ref.\cite{smna} is improper and the periodicity discussed 
in Ref.\cite{hagen} is of no significance. 

In the comment \cite{hagen} it was claimed that factors of $(-1)^{2l}$
are omitted in the expression of the two-particle partition function Eq.(26)
of Ref.\cite{smna} or in that of the second virial coefficient. 
(Here $l$ is the isospin quantum number of the particle.) However, 
these factors are unnecessary unless we impose a relationship, between 
the (intrinsic) statistics of the particles and their isospins, 
which would be similar to the spin-statistics relation in (3+1) dimensions. 
Obviously such a relation does not exist. 

The NACS particles are described in Ref.\cite{smna} as point like spinless 
sources which interact with each other through the non-Abelian Chern-Simons 
gauge fields. Thus, when we describe the NACS particles in the regular gauges   
such as the Coulomb gauge or the holomorphic gauge, the wave function for the 
many NACS particle system is symmetric under an exchange of any pair of the 
particles. It was also claimed in Ref.\cite{hagen} that the configuration 
space wave functions for all states are taken to be symmetric in 
Ref.\cite{smna} and the resultant expression for the second virial 
coefficient reduces to the bosonic one as $\kappa \rightarrow \infty$.  
Contrary to the claim, the configuration space wave function is symmetric 
only when its isospin counterpart is symmetric; it is antisymmetric when
its counterpart is antisymmetric.
It becomes manifest, as we rewrite the two-particle partition function 
Eq.(21b) of Ref.\cite{smna} as follows
\begin{eqnarray}
Z_2^\prime &=& \int d^2 z \sum_{m_1,m_2}\Bigl[<(m_1,m_2)|\otimes 
<z_S|e^{-\beta H_{\rm rel}}\nonumber\\ & &\quad  
|z_S> \otimes|(m_1,m_2)>+ <[m_1,m_2]|\otimes <z_A|e^{-\beta H_{\rm rel}}
\nonumber\\& &\quad 
|z_A> \otimes|[m_1,m_2]>\Bigr]\label{2part}
\end{eqnarray}
where $|(m_1,m_2)>$ ($|[m_1,m_2]>$) denotes the symmetric (antisymmetric) 
isospin state and $|z_S>$ ($|z_S>$) denotes the symmetric (antisymetric) 
configuration state under the exchange of the particle 1 and 
particle 2. This expression for the two-particle partition function 
Eq.(21b) in Ref.\cite{smna} yields that the second virial coefficient 
reduces to 
\begin{eqnarray}
\Delta B_2(l, T) &= & \frac{1}{(2l+1)^2} \sum^{2l}_{j=0} (2j+1)
\Biggl[\frac{1+(-1)^j}{2}B^B(T)
\nonumber\\ & & \qquad
+ \frac{1-(-1)^j}{2}B^F(T) \Biggr]
\end{eqnarray}
if the non-Abelian statistical interaction is switched off. 
It is not certainly the bosonic result. 
Here it is clear that the states with $j={\rm even\,\,\, integer}$ contribute
to the virial coefficient as free bosonic states while those with
$j={\rm odd\,\,\, integer}$ as free fermionic states.  
Note that the limit $\kappa\rightarrow \infty$ does not coincide with
the case where the statistical interaction is switched off. It implies
that the second virial coefficient contains some non-perturbative effect.

The parameter $\alpha$ in Ref.\cite{hagen} is written by $1/(8\pi\kappa)$ 
in terms of the Chern-Simons coefficient $\kappa$. Observing that 
$\omega_j$ is an integer multiple of $\alpha$, Hagen also asserted that
there exists a periodicity just as in the Abelian theory: $\alpha  
\rightarrow \alpha+2$. However, if we translate $\alpha$ by 2, the theory
does not exist any more, since $1/(2\alpha+4)$ and $1/(2\alpha)=4\pi\kappa$ 
cannot be integral at the same time. (Recall that $4\pi\kappa$ must be 
an integer for the theory to be consistent.)
Thus, the asserted periodicity is of no significance. 

Although the author mostly does not agree with Hagen\cite{hagen}, 
he must admit that a mistake was made in Ref.\cite{smna}. 
As pointed out in Ref.\cite{hagen}, no differentiation was made between 
even and odd $[\omega_j]$. If this point is taken into account, 
the second virial coefficient takes the same
expression Eq.(30) of Ref.\cite{smna}. But $\delta_j$ should read now
as $\delta_j=\omega_j-2[\omega_j/2]$.

\acknowledgements{
This work was supported in part by nondirected research fund, Korea Research 
Foundation 1994 and by KOSEF through the Center for Theoretical
Physics at Seoul National University.}

\end{document}